\documentclass{aa}
\usepackage{graphicx}
\begin{document}


%

   \title{Size determination of the Centaur Chariklo from
millimeter-wavelength bolometer observations}

   \author{W. J. Altenhoff, K. M. Menten, \and F. Bertoldi
          }

   \offprints{W. J. Altenhoff}

   \institute{Max-Planck-Institut f\"ur Radioastronomie,
              Auf dem H\"ugel 69, D--53121 Bonn, Germany\\
              \email{waltenhoff, kmenten, fbertoldi@mpifr-bonn.mpg.de}
              }  

   \titlerunning{Size of Chariklo}
   \date{Received; accepted}


\abstract{ Using the Max-Planck Millimeter Bolometer Array (MAMBO) at the
IRAM 30m telescope we detected emission
at 250 GHz from the Centaur Chariklo (1997~CU26). The observed 
continuum flux density 
implies a photometric diameter of 273 km. The resulting geometric
albedo is 0.055, somewhat higher than expected from a comparison
with most of the other few Centaurs and cometary nuclei for
which such data are available. 
              }
 
\maketitle
 
%
 
\section{Introduction}

Within the last decade, our view of the outer solar system has been
revolutionized by the discovery of a large number of
``trans-Neptunian'' objects, which appear to be remnant planetesimals
from the formation of the solar system (Jewitt 1999).  These so-called
(Edgeworth-) Kuiper Belt objects (EKOs, KBOs, or TNOs) have orbits in
the range from 30 to 50 AU (see Jewitt \&\ Luu 2000 for a review).
Near the Kuiper Belt, two other populations of asteroids were
detected recently, the Centaurs (named by Kowal 1989 after his
detection of Chiron), which have perihelia between Jupiter and
Neptune, and the scattered disk objects (SDO, or SKDO), having
perihelia near 35 AU, semimajor axes near 85 AU, high eccentricities,
and intermediate inclinations (Chadwick et al. 2000).

Systematic surveys for these objects allow estimates of their total
populations. 
Based on their survey, Jewitt et al. (1996) deduce a total number of
70000 EKOs with diameters, $d \ge$100 km and of 2600
Centaurs with $d\ge$70 km. Chadwick et al. (2000) 
estimate that there are  31000 SDOs with $d\ge$100 km. 
According to Parker (2000), presently 330 EKOs and 55 Centaurs
(including 5 SDOs) are known with their orbital elements; during the
last few years the numbers of catalogued EKOs and Centaurs have grown
exponentially, more than doubling each year. 
The obvious reasons why
these distant asteroids have been detected only recently are their
``small'' sizes and low albedos, resulting in faint optical emission
($m_V \sim 19$--24) and very low infrared (IR) and radio flux
densities. The resulting lack of precise thermal (i.e. infrared-,
(sub)millimeter-, or radio-wavelength) measurements in turn precludes
accurate albedo and size determinations.

These distant asteroids (and comets) are of great interest because
they are likely to consist of pristine and mostly unaltered matter
from the solar nebula. Their nearly black surface is thought to be the
result of cosmic particle bombardment (e.g., Johnson et al. 1987).
The KBOs move in a dynamically stable region, while the Centaurs orbit
in a dynamically unstable range between the major gas planets, which
limits their dynamical life times to about $10^6$ years. The
unexpectedly large population of Centaurs can only be maintained by
continuous replenishment from the reservoir of EKOs (Stern \&\ Campins
1996). Similarly, the SDOs are thought to be ejected from the Kuiper Belt
by Neptune scattering (Chadwick et al. 2000).

Circumstantial evidence for a Kuiper belt origin of the Centaurs comes
from the facts that both, EKOs and Centaurs, have similar reflectance
spectra, cover probably a similar size range, and probably have
similar albedos (see below). Numerical orbit integrations by Levison
\&\ Duncan (1997) indicate that at the end of its lifetime a Neptune
encountering body such as a Centaur may become a Jupiter family short
period ``comet''. Indeed, attempts to subdivide these distant icy
bodies into asteroids on the one hand and comets one the other seem
artificial. For example, Chiron shows many of the
characteristics of a comet and yet is the prototypical Centaur, while
the bona fide comets P/Oterna and P/Schwassmann-Wachmann I are listed
by Jewitt \&\ Kalas (1998) and Jewitt \&\ Luu (2000) as Centaurs.

Optical wavelength reflection spectra are of fundamental importance
for understanding the nature of these bodies.  From their observations of
five KBOs, Jewitt \&\ Luu (1998) found a linear relationship between the
$V$--$J$ color index and the absolute red magnitude $M_R$.  Tegler \&\
Romanishin (1998) observed a few KBOs and Centaurs, finding a bimodal
distribution in the $B$--$V$, $V$--$R$ color diagram for both types of
asteroids.  Davies et al. (2000) obtained photometry for a larger
sample of KBOs.  While they confirm previous measurements for those
objects of their sample that had been studied before, their results in
total make the existence a color-distance correlation and a bimodal
distribution of KBOs appear less convincing.

Using optical data alone, the sizes of these distant bodies can only
be estimated from the observed optical magnitude by assuming a
geometric ($V$-band) albedo $p_V$. Jewitt \&\ Luu (1995)
used for their population statistics a ``comet-like geometric albedo
of 0.04'', but it is still unknown how good an assumption this is,
considering that the Pluto-Charon binary, frequently classified as a
KBO, has a geometric albedo around 0.60.

Given these uncertainties, the derived diameters can be wrong
by more than a factor of 2.  This is an unfortunate situation, since an
accurate size determination is an indispensable prerequisite for any
physical interpretation, in particular for deriving surface properties
and the mass of a single object and, thus, the total mass contained in
all the distant asteroids. There is obviously an urgent need for good 
size and albedo determinations.

Earlier observations with the IRAM 30 m telescope have shown that
precise mean size determinations of asteroids can be made using
careful photometry near a frequency of 250 GHz (1.2 mm wavelength;
Altenhoff et al. 1994; Altenhoff \&\ Stumpff 1995).  
Altenhoff et al. (1996) showed that
the quality of a mm-wavelength photometric size determination can match that of
occultation measurements, if the mm observations cover the rotation
period of the asteroid.

Given their large distances and small diameters EKOs are expected to be
very faint at millimeter wavelengths. Expected 250 GHz flux densities
for the largest of these objects ($d \approx 600$ km) at a typical
distance of 40 AU are around 0.5 mJy. Given this, their
detection would require prohibitive amounts of observing time even
with the largest millimeter-wavelength telescopes
equipped with sensitive bolometer detectors.  However, due to their
proximity, the flux densities of some Centaurs should be higher
and observing them would yield results that are relevant to KBOs as well.
For example, the signal of Chariklo, formerly known as 1997~CU26, is
expected to be five times stronger than the above value and thereby
detectable.  This Centaur was detected with
the Spacewatch telescope in 1997 (Scotti 1997).  It has an absolute
magnitude, $H$ (i.e. visual magnitude, reduced to 1 AU geocentric and
heliocentric distance and phase angle 0; see, e.g., Bowell et al. 1989) 
of 6.4, corresponding to an estimated diameter of 353 km, when assuming an 
albedo of 0.04.  It is the largest known Centaur and has a very dark surface,
supposedly comparable to that of Pholus (Mueller et al. 1992).
 
Mid-infrared photometry at 20 $\mu$m by Jewitt \&\ Kalas (1998)
resulted in an improved diameter of 302 km. This value depends on
assumptions of the asteroidal model, the IR emissivity, IR phase, the
``beaming parameter'', etc., which are not all accurately known. This can
introduce systematic uncertainties in the size estimate that are bigger
than the quoted error of 30 km. In contrast, we know from experience
that at 250 GHz the emissivity, $e$,  and the beaming parameter are close to
unity and that the phase effect is negligible 
(for discussions of these quantities 
see Spencer et al. 1989; Altenhoff et al. 1994; Jewitt \&\ Kalas 1998).
Thus (sub)millimeter observations in principle are capable of yielding
more accurate diameter estimates than IR data.  In the following we
describe our 250 GHz observations of Chariklo.

\section{Observations}

Our observations were made with the IRAM 30m telescope on Pico Veleta,
Spain, using a 37 channel version of the {\bf Ma}x-Planck {\bf
M}illimeter {\bf Bo}lometer (MAMBO), built at the Max-Planck-Institut
f\"ur Radioastronomie (Kreysa et al.  1998). This
is a hexagonally close packed, diffraction limited bolometer 
array operating at
0.3 K. A common filter defines a bandpass centered at 250 GHz, if
folded with black body spectra corresponding to minor planets.
The beamwidth is $11''$ (FWHM).
Observing modes included drift scans for pointing and calibration
measurements and an ON--OFF technique described by Bertoldi et
al. (2000), which provides high (sub-mJy) sensitivity by employing sky
noise subtraction.

The flux density scale was derived from observations of 
asteroid 1 Ceres (see Altenhoff et al. 1996)
and the planets Mars and Uranus from which a
calibration factor of 12500 counts per Jy was derived.

The calculation of the ephemerides for the minor planets 1 Ceres and
10199 Chariklo was based on the orbital elements, published by Shor
(1999) and Marsden and Williams (1999), respectively, adjusting the
osculating epoch to the date of observations by taking into account
the perturbations by the major planets. Since the orbital elements for
Chariklo were determined from 344 observations covering 11(!) 
years (including prediscovery data), the
resulting position accuracy is $\sim 0.5''$ and thus easily
sufficient for our observations.

Typical rotation periods of asteroids and comets are between 5 and 15
hours (see, e.g., light curve parameters given by Shor 1999).  Since
Chariklo's rotation period is unknown, we observed it 
during six short periods on different days in order to cover
different intervals of a possibly existing ``light curve''.  Averaging
the data from these individual observations will yield a mean
diameter. Given that the expected signal was only $\approx 2.5$ mJy
precluded an actual measurement of light variations with plausible
amplitudes or obtaining constraints on the three-dimensional shape of
this Centaur.

We observed Chariklo between 1999 December 9 and 2000 February 29 at
night time under good weather conditions. Its mean geocentric
distance, $\Delta$, was 12.62 AU and its mean heliocentric distance,
$r$, 13.35 AU.  The data were reduced using an improved version of the
MOPSI package developed by R. Zylka, allowing for effective sky noise
subtraction and calibration, resulting in a higher sensitivity than
attainable with previous data reduction procedures.

Our observations yield a mean 250 GHz flux density of $S_\nu = 2.08
\pm0.30$ mJy after a total integration time of 2.4 hours; Fig.~1
shows the signals of the bolometer channel centered on Chariklo and
the other bolometer channel (showing noise) as a function of the
integration time. 


\section{Results and discussion}

Assuming a diameter of 302 km and a geometric albedo of 0.045, as derived
by Jewitt \&\ Kalas (1998), we predict a flux density of 2.54 mJy using
the equations given by Altenhoff et al. (1994).
Reversing the argument, we can use above geometric albedo (which we assume to
be equal to the Bond albedo $A$; see Jewitt \&\ Kalas 1998) 
to calculate Chariklo's equilibrium 
temperature $T_{\rm eq}$ from
$
T_{\rm eq} = f (1-A)^{0.25} r^{-0.5},
$
where $f$ = 277 K for ``fast
rotation'', a condition met by almost all asteroids, 
and $r$ is measured in AU. The diameter, $d$, is determined from
$
S_\nu = {{\pi d^2}\over{4 \Delta}} e B_\nu(T_{\rm B}),
$
where $e = 1$ (see above), $T_{\rm B}$ is the brightness temperature
above the cosmic microwave background ($T_{\rm B} = T_{\rm eq} - 2.73$ K),
and $B_\nu(T)$ is the Planck function. With 
$S_\nu = 2.08\pm0.30$ mJy we obtain $d = 273 \pm 19$ km.  
Plugging this newly derived
diameter into eq. (1) of Jewitt \&\ Kalas (1998), yields a new value
of $0.055\pm0.008$ for the geometric albedo.  Our assumption of fast
rotation should -- at millimeter wavelengths -- apply for all rotation
periods under 50 hours. It appears to be supported by the results of
Peixinho et al. (2000), who tried to determine Chariklo's optical
light curve, finding a rotation period of either 15 or 34 h.

Assuming no rotation or, equivalently, that the asteroidal
rotation axis is pointing toward the observer, one calculates for the
sun-lit surface an equilibrium temperature that is 19\%\ higher than
the fast rotation value, resulting in a 9\% smaller derived diameter.
Applying this to our numbers yields $d$ = 250 km and $p_V = 0.065$ for
Chariklo.  Since the surface depth probed by thermal emission is
approximately proportional to the wavelength, the time scale
discriminating between the fast or slow rotation scenarios is also
wavelength dependent. Thus, while our adoption of fast rotation
appears to be appropriate for mm wavelengths, the no rotation
assumption of Jewitt \&\ Kalas (1998) may be applicable to their $20
\mu$m data.

\begin{table}
      \caption[]{Diameters and albedos of Centaurs and cometary nuclei}                 \label{diatab}                               
      \[
\begin{tabular}{lcllcc}
                    \hline
            \noalign{\smallskip}

Object &   Band/ &  Diameter & $p_V^{\mathrm{a}}$  &      Ref.\\
       &   Method&   [km]    & (\%)    \\

            \noalign{\smallskip}
            \hline
            \noalign{\smallskip}

Chariklo       & $20 \mu$m      & 302   & \,~$4.5 \pm 1.0$       & 1   \\
               &  1.2 mm        & 275   & \,~$5.5 \pm 0.5$   &     2 \\
Chiron         & $10/20 \mu$m   & 180   & $14^{+6}_{-3}$      & 3   \\
               &  1.2 mm        & 168   & $13^{+4}_{-3}$      & 4   \\
               & occult.$^{\mathrm{b}}$    & 179.2 & $14.8\pm 2.2$       & 5   \\ 
Pholus         & $20 \mu$m      & 189   & \,~$4.4\pm 1.3$       & 6   \\
P/Halley       & Giotto$^{\mathrm{c}}$     &  \,~10   & \,~$3.9\pm 0.3$       & 7,8 \\
P/Arend-Rigaux & 5--$20 \mu$m      &  \,~10  &\,~$2.8\pm 0.5$       &9  \\
P/Neujmin 1    & $10/20 \mu$m      &  \,~20   &\,~$2.5\pm 0.5$       &10   \\
P/Tempel 2     & $20 \mu$m      &  \,~11.8 & ~$\,2.2^{+0.4}_{-0.6}$& 11 \\

            \noalign{\smallskip}
            \hline
\end{tabular}
      \]
\begin{list}{}{}

\item[$^{\mathrm{a}}$] Visual geometric or albedo
\item[$^{\mathrm{b}}$] Diameter determination using stellar occultation
\item[$^{\mathrm{c}}$] Diameter determination using data from the Giotto
spacecraft
\end{list}
References: 
(1) Jewitt \&\ Kalas (1998), (2) this paper, (3) Campins et al. (1994), (4) Altenhoff \&\
Stumpff (1995), (5) Bus et al. (1996), (6) Davies et al. (1993), 
(7) Jewitt et al. (1982), albedo derived from diameter in ref. 8, 
(8) Keller et al. (1987),
(9) Millis et al. (1988), (10) Campins et al. (1987),
(11) A'Hearn et al. (1989)
   \end{table}

Above albedo values for Chariklo are similar to 0.04, the value often
assumed for EKOs and related objects.  Table \ref{diatab} presents a
critically selected collection of the few meaningful diameter and
albedo determinations of Centaurs and cometary nuclei that are
available in the literature. As discussed above, Centaurs and the
short period Jupiter comets are likely to originate from the Kuiper
belt.
Our millimeter data for Chariklo, 
confirm the earlier IR data  for this object.

The derived albedo values vary from 0.02 to possibly 0.14, but the
median value is still near 0.04. We note Chiron's high albedo,
meticulously derived by Campins et al. (1994), who subtracted the coma
contribution.  Whether this high geometric albedo is unique, is
presently difficult to answer. The Pluto-Charon binary with its albedo
of $\sim0.6$ (Marcialis et al. 1992) may have a different structure
than typical EKOs.  We further note that C/Hale-Bopp may have a high
albedo around 0.3. This estimate is based on this comet's unexpectedly small
nuclear diameter indicated by the millimeter observations of
Altenhoff et al. (1999) and its absolute magnitude, $H$,
of $\approx 8.7$, which is derived 
from the pre-discovery observations of McNaught
(1995).  However, Hale-Bopp's albedo determinations have to be viewed
with caution since this object showed cometary activity already at a
distance $\Delta$ $\ge$13 AU. It is conceivable that the higher
albedos of active Centaurs/comets are caused by the partial loss of
their dark surface covers.

While the similar albedos of Centaurs and cometary nuclei may
indicate a relation between both classes of objects,  
these groups have quite different reflection spectra and size distributions.
The latter point is hinted at by the data compiled in Table~\ref{diatab},
although the statistics are poor.
Moreover, according to Luu \&\
Jewitt (1996) most of the Centaurs differ from cometary nuclei and
Trojan asteroids by their extreme red color and IR absorption lines.
Recently, Tegler \&\ Romanishin (2000) reported that most of the
observed KBOs show extremely red colors as well.  
Thus there seem to be marked
differences between KBOs and Centaurs on the one side and cometary nuclei
and Trojans on the other.
Nevertheless, the above does not necessarily exclude the
possibility that the Jupiter family comets originate in the Kuiper
belt. However, if that scenario were true, the data seem to imply
that the sizes and surfaces of these objects are 
substantially altered during their way to the Sun.


\begin{figure}
   \centering
   \includegraphics[width=0.40\textwidth]{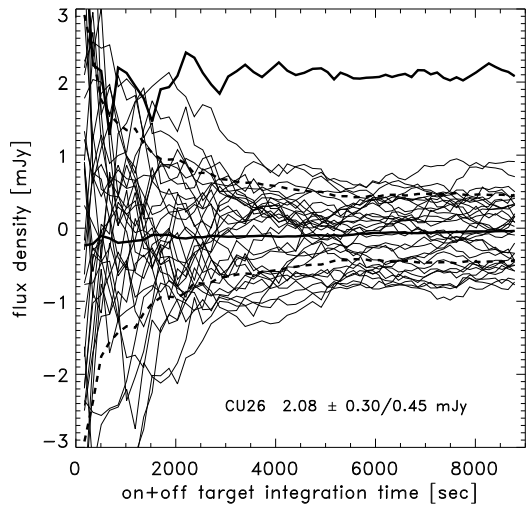}
    \caption[]{Observed signal of Centaur Chariklo.  Plotted is the
       integrated signal for all bolometer channels as function of
       integration time.  Channel 1 (on source) is plotted as heavy line,
       showing a signal of 2.08 mJy. The dashed lines indicate the rms
       dispersion of all off-source channels, which is 0.45 mJy.  
             }
   \label{FigGam}%
\end{figure}

\section{Conclusions and Outlook}

Our millimeter wavelength detection of the Centaur Chariklo
yields an unbiased size determination and an albedo estimate for
this object.
Since Centaurs are very likely to be
Edgeworth-Kuiper-Objects that have moved inwards from the Kuiper Belt,
our results may also have implications for bona fide EKOs.  At
distances around 40 AU, these will be very difficult to detect with
today's millimeter-wavelength instrumentation. However, with next
generation facilities such as the Atacama Large Millimeter Array
(ALMA) it will not only be possible to detect hundreds of EKOs.
ALMA will also allow interferometric, {\it direct}
size measurements of such objects (see, e.g., Menten 2000).   


\begin{acknowledgements}
        We would like to thank Ernst Kreysa and the MPIfR bolometer group,
Robert Zylka for providing an improved version of his MOPSI data 
reduction package, Clemens Thum for flexibly scheduling this 
project as required by its need for optimum observing conditions, and
the IRAM staff at the 30m telescope for their able technical support.
\end{acknowledgements}


\begin{thebibliography}{}

\bibitem[1989]{A89} A'Hearn, M. F., Campins, H., Schleicher, D. G.,
   Millis, R. L., 1989, ApJ 347, 1155

\bibitem[1996]{A96} Altenhoff, W. J., Baars, J. W. M., Schraml, J. B., 
   Stumpff, P., von Kap-herr, A., 1996, A\&A 309, 953

\bibitem[1994]{A94} Altenhoff, W. J., Johnston, K. J., Stumpff, P., 
   Webster, W. J., 1994, A\&A 287, 641

\bibitem[1995]{A95} Altenhoff, W. J., Stumpff, P., 1995, A\&A 293, L41

\bibitem[1999]{A99} Altenhoff, W.J., et al., 1999, A\&A 348, 1020 


\bibitem[2000]{B00} Bertoldi, F., et al.,  2000, A\&A 360, 92

\bibitem[1989]{B99} Bowell, E., Hapke, B., Domingue, D., Lumme, K.,
   Peltoniemi, J., Harris, A. W., 1989, in: R. P. Binzel, T. Gehrels, 
   M. S. Shapley (eds.), Asteroids II. Univ. Arizona Press, Tucson, p. 524

\bibitem[1996]{B96} Bus, S. J., et al., 1996, Icarus 123, 478

\bibitem[1987]{C87} Campins, H., A'Hearn, M.F., McFadden, L-A., 1987,
   ApJ 316, 847

\bibitem[1994]{C94} Campins, H., Telesco, C. M., Osip, D. J., Rieke,
   G. H., Rieke, M. J., Schulz, B., 1994, AJ, 108, 2318

\bibitem[2000]{C00} Chadwick, A., Trujillo, C., Jewitt, D.C., Luu, J.X.,
    2000, ApJ 529, L103

\bibitem[1993]{D93} Davies, J., Spencer, J., Sykes, M., Tholen, D.,
   Green, S. 1993, IAU Circ. 5698 


\bibitem[2000]{D00} Davies, J. K., et al., 2000, Icarus 146, 253

\bibitem[1999]{J99} Jewitt, D., 1999, Annu. Rev. Earth Planet. Sci. 27, 287

\bibitem[1998]{J98a} Jewitt, D., Kalas, P., 1998, ApJ 499 L103


\bibitem[1995]{J95} Jewitt, D., Luu, J. X., 1995, AJ 109, 1867

\bibitem[1998]{J98} Jewitt, D., Luu, J., 1998, AJ 115, 1667

\bibitem[1996]{J96} Jewitt, D., Luu, J., Chen, J., 1996, AJ 112, 1225

\bibitem[2000]{J00} Jewitt, D. C., Luu, J. X., 2000,
   in: V. Mannings, A. P. Boss, S. S. Russell (eds.), 
   Protostars and Planets IV. Univ. Arizona Press, Tucson, p. 1201

\bibitem[1982]{J82} Jewitt, D.C., et al., 1982, IAU Circular 3737

\bibitem[1987]{J87} Johnson, R. E., Cooper, J. F., Lanzerotti, L. J.,
   Strazzulla, G., 1987,
  A\&A 187, 889

\bibitem[1987]{K87} Keller, H. U., Delamare, W. A., Huebner, W. F. et
      al., 1987, A\&A 187, 807

\bibitem[1989]{K89} Kowal, C.T., 1989, Icarus 77, 118


\bibitem[1998]{K98} Kreysa, E., et al., 1998, in: T. G. Phillips (ed.),
   Advanced Technology MMW, Radio, and Terahertz Telescopes, 
   Proc. SPIE 3357, 319

\bibitem[1997]{Le97} Levison, H.F., Duncan, M.J., 1997, Icarus 127,13

\bibitem[1996]{L96} Luu, J., Jewitt, D., 1996, AJ 112, 2310

\bibitem[1997]{L97} Luu, J., 
1997, Nature 387, 573

\bibitem[1992]{M97} Marcialis, R. L., Lebofsky, L. A., DiSanti, M. A., 
    Fink, U., Tedesco, E. F., Africano, J., 1992, AJ 103, 1389

\bibitem[1999]{M99} Marsden, B. G., Williams, G. V., 1999,
  Minor Planet Circular No. 33905

\bibitem[1995]{M95} McNaught, R.H., 1995, IAU Circular 6198

\bibitem[2000]{M00} Menten, K. M., 2000, in: J. Bergeron, A. Renzini (eds.), 
   ESO Astrophysics Symposia, From Extrasolar Planets to Cosmology: 
   The VLT Opening Symposium. Springer, Berlin, p. 78 

\bibitem[1988]{M88} Millis, R. L., A'Hearn, M. F., Campins, H., 1988,
   ApJ 324, 1194 

\bibitem[2000]{P00} Parker, J. W., 2000,
  The Kuiper Belt Electronic Newsletter Nr. 12

\bibitem[2000]{Pe00} Peixinho, N., Lacerda, P., Roos-Serote, M., Ortiz, J.,
    Doressoundiram, A., 2000, BAAS 32, 961

\bibitem[1997]{S97} Scotti, J. V., 1997, Minor Planet Circular 1997-D11

\bibitem[1999]{S99} Shor, V. A. (editor), 1999, Ephemerides of Minor Planets, 
     St. Petersburg

\bibitem[1989]{S89} Spencer, J. R., Lebofsky, L., Sykes, M. V., 1989, Icarus
     78, 337

\bibitem[1996]{S96} Stern, A., Campins, H., 1996, Nature 382, 507

\bibitem[1998]{TR98} Tegler, S. C., Romanishin, W., 1998, Nature 392, 49

\bibitem[2000]{TR00} Tegler, S. C., Romanishin, W., 2000, Nature 407, 979

\end{thebibliography}
\end{document}